\documentclass[11pt]{article}

\usepackage[cmtip,arrow]{xy}
\usepackage{pb-diagram,pb-xy}

\newlength{\vshift}
\newlength{\hshift}
\usepackage{amssymb,amsopn}

\def\al{\alpha}

\def\ds{\stackrel{\star}{,}}

\def\p{\partial}

\def\lb{\lbrack}
\def\rb{\rbrack}
\def\p{\partial}

\begin{document}

\begin{titlepage}
\rightline{LMU-ASC 09/06}
\rightline{MPP-2006-17}

\vspace{2em}
\begin{center}

{\huge{\bf Twisted Gauge Theories}}

\vskip 3em

{{\sc \bf Paolo Aschieri${}^{5}$, Marija Dimitrijevi\' c${}^{4}$, \\
Frank Meyer${}^{1,2}$, Stefan Schraml${}^{2}$ and Julius Wess${}^{1,2,3}$ }}

\vskip 2em

${}^{1}$Arnold Sommerfeld Center for Theoretical Physics\\
Universit\"at M\"unchen, Fakult\"at f\"ur Physik\\
Theresienstr.\ 37, 80333 M\"unchen, Germany\\[1em]

${}^{2}$Max-Planck-Institut f\"ur Physik\\
F\"ohringer Ring 6, 80805 M\"unchen, Germany\\[1em]

${}^{3}$Zentrum f\" ur Mathematische Physik \\ DESY und Universit\" at Hamburg\\Notkestr. 85, 22607 Hamburg, Germany \\[1em]

${}^{4}$University of Belgrade, Faculty of Physics\\
Studentski trg 12, 11000 Beograd, Serbia and Montenegro\\[1em]

${}^{5}$Dipartimento di Scienze e Tecnologie Avanzate\\
Universit\' a del Piemonte Orientale, and INFN\\
Via Bellini 25/G 15100 Alessandria, Italy\\[2em]

\end{center}


\begin{abstract}

Gauge theories on a space-time that is deformed by the Moyal-Weyl product are constructed by twisting the coproduct 
for gauge transformations.  This way a
deformed Leibniz rule is obtained, which is used to construct gauge invariant quantities. 
The connection will be enveloping algebra valued in a particular representation of the Lie algebra. 
This gives rise to additional fields, which couple only weakly via the deformation parameter $\theta$ and 
reduce in the commutative limit to free fields.   
Consistent field equations that lead to conservation laws are derived and some properties of such theories are
discussed. 
\end{abstract}
\vspace*{0.5cm}



\vspace*{0.5cm}
\quad\scriptsize{email: aschieri,meyerf,wess@theorie.physik.uni-muenchen.de, dmarija@phy.bg.ac.yu,} \\
\vspace*{0.5cm}
\qquad \qquad \qquad\scriptsize{schraml@mppmu.mpg.de}
\vfill

\end{titlepage}\vskip.2cm

\newpage
\setcounter{page}{1}
\newcommand{\Section}[1]{\setcounter{equation}{0}\section{#1}}
\renewcommand{\theequation}{\arabic{section}.\arabic{equation}}

\Section{Introduction}
The aim of this work is to construct and investigate gauge theories on deformed space-time structures that
are defined by an associative but noncommutative product of $C^\infty$ functions. Such products are known as
star products; the best known is the Moyal-Weyl product \cite{Weyl:1927vd,Moyal:1949sk}. In this letter we 
shall deal with this product exclusively. 

From previous work  \cite{Aschieri:2005yw,Aschieri:2005zs,Meyer:2005as} we know that the usual algebra of 
functions and the algebra of vector
fields can be represented by differential operators on the deformed manifold. The deformed diffeomorphisms
have been used to construct a deformed theory of gravity. Here we shall show that along the same lines a
deformed gauge theory can be constructed as well. The algebra, based on a Lie algebra, will not change but
the comultiplication rule will. This leads to a deformed Hopf algebra. In turn this gives rise to deformed gauge
theories because the construction of a gauge theory involves the Leibniz rule that is based on the
comultiplication. 

Covariant derivatives can be constructed by a connection. Different to a 
usual gauge theory the connection cannot be Lie algebra valued.  
The construction of covariant tensor fields (curvature or field strength)
and of an invariant Lagrangian is completely analogue to the undeformed case. 
Field equations can be derived
and it can be shown that they are consistent. This leads to conserved currents. It is for the first time
that it is seen that deformed symmetries also lead to conservation laws; note that the Noether theorem is not
directly applicable in the noncommutative context. 

The deformed gauge theory has interesting new features. 
We start with a Lie$(G)$-valued connection and show that twisted 
gauge transformations close in Lie($G$), however consistency of the equation 
of motion requires the introduction of additional, new vector potentials.
The number of these extra vector potentials is representation dependent but 
remains finite for finite dimensional representations. 
Concerning the interaction, the Lie algebra valued fields and the 
new vector fields behave quite differently. 
The interaction of the Lie algebra valued fields can be seen 
as a deformation of the usual
gauge interactions; for vanishing deformation parameters the interaction 
will be the interaction of a usual gauge theory. The interactions of the new 
fields are deformations of a free field theory for vector potentials;
for vanishing deformation parameters the fields become free. 
As the deformation parameters are supposed to be
very small we conclude that the new fields are practically dark with respect 
to the usual gauge interactions.

Finally we discuss the example of a $SU(2)$ gauge group in the two 
dimensional representation.

The treatment introduced here can be compared with previous ones. In \cite{Seiberg:1999vs}
the noncommutative  gauge transformations for $U(N)$ have an undeformed comultiplication.  The action is the same as in (\ref{Sg}) 
if we restrict our discussion, valid for any compact Lie group, to $U(N)$ in the $n$-dimensional matrix
representation. In other terms we show that noncommutative $U(N)$ gauge theories have usual 
noncommutative gauge transformations and also twisted gauge transformations.
In \cite{Jurco:2000fb,Jurco:2000fs,Jurco:2001my,Jurco:2000ja,Jurco:2001rq} the situation is different because 
we consider field dependent transformation parameters. 


\Section{Algebraic formulation}

A noncommutative coordinate space can be realized with the help of the Moyal-Weyl product \cite{Weyl:1927vd,Moyal:1949sk}. On such a space we are going to construct gauge theories based on a Lie algebra. 

We start from the linear space of $C^{\infty}$ functions on a smooth manifold ${\cal M}$, $Fun({\cal
M})$. To define an algebra $\mathcal{A}_\theta$ we shall use the associative but noncommutative Moyal-Weyl product. 
The algebra defined with the usual, commutative point-wise product we refer to as 
the  algebra of $C^{\infty}$ functions. 

The Moyal-Weyl product is defined as follows 
\begin{eqnarray}
f,g \hspace*{-2mm}&{\in}&\hspace*{-2mm} Fun({\cal M})\nonumber\\
f\star g \hspace*{-2mm}&=&\hspace*{-2mm} \mu \{ e^{\frac{i}{2}\theta^{\rho\sigma}\p_\rho \otimes\p_\sigma}f\otimes g \}
\label{1}\\
\mu\{f\otimes g\} \hspace*{-2mm}&= &\hspace*{-2mm} f\cdot g , \nonumber
\end{eqnarray}
where $\theta^{\rho\sigma}=-\theta^{\sigma\rho}$ is $x$-independent.
The $\star$-product of two functions is a function again
\begin{eqnarray}
\mu_\star:\quad\quad Fun({\cal M})\otimes Fun({\cal M}) 
\hspace*{-2mm}&\to &\hspace*{-2mm} Fun({\cal M}) ,\nonumber\\
\mu_\star\{f\otimes g\} \hspace*{-2mm}& = &\hspace*{-2mm} f\star g .\label{2}
\end{eqnarray}
Derivatives are linear maps on $Fun({\cal M})$
\begin{eqnarray}
\partial _\rho:\quad\quad Fun({\cal M}) 
\hspace*{-2mm}&\to &\hspace*{-2mm} Fun({\cal M}) ,\nonumber\\
f \hspace*{-2mm}&\mapsto &\hspace*{-2mm} \partial_\rho f .\label{3}
\end{eqnarray}
The Leibniz rule extends these maps to the usual algebra of $C^{\infty}$ functions
\begin{equation}
(\p_\rho (f\cdot g)) = (\p_\rho f)\cdot g + f\cdot (\p_\rho g) .\label{4}
\end{equation}
This concept can be lifted to the algebra ${\cal{A}}_\theta$ \cite{Wess:2003da}
\begin{eqnarray}
{\partial}^\star_\rho: \quad\quad f \hspace*{-2mm}&\mapsto &\hspace*{-2mm} 
{\partial}^\star_\rho f \equiv \partial _\rho f \nonumber\\
&&\hspace*{-1cm}\p^\star_\rho (f\star g) = (\p^\star_\rho f)\star g 
+ f\star (\p^\star_\rho g) .\label{4'}
\end{eqnarray}
The last line is true because $\theta^{\mu\nu}$ is $x$-independent.

Analogously to differential operators acting on the usual algebra of functions we define differential operators on ${\cal {A}}_\theta$
\begin{equation}
{\cal{D}}^\star \star f = \sum_{n}d^{\rho_1\dots \rho_n}\star\p^\star_{\rho_1}\dots\p^\star_{\rho_n} f .\label{5}
\end{equation}
This is well defined, $\star$ and $\p^\star_\rho$ always act on functions. The product of such differential operators can be computed with the help of the Leibniz rule.

We can now define the $\star$-product as the action of a bilinear differential operator
\begin{equation}
f\star g = \mu \{ {\cal F}^{-1}\, f\otimes g \} ,\label{6}
\end{equation}
with 
\begin{equation}
{\cal F}^{-1} = e^{\frac{i}{2}\theta^{\rho\sigma}\p_\rho \otimes\p_\sigma} .\label{7}
\end{equation}
This differential operator can be inverted
\begin{equation}
f\cdot g = \mu_\star \{ {\cal F}\, f\otimes g\} .\label{8}
\end{equation}
Equation (\ref{8}) can also be written in the form \cite{Aschieri:2005yw}
\begin{equation}
f\cdot g = \Big( \sum _{n=0}^\infty \Big(-\frac{i}{2}\Big)^n
\frac{1}{n!}\theta^{\rho_1\sigma_1}\dots
\theta^{\rho_n\sigma_n}\Big(\partial_{\rho_1}\dots\partial_{\rho_n}f\Big) \star
\partial^\star_{\sigma_1}\dots\partial^\star_{\sigma_n}\Big) \star g .\label{9}
\end{equation}
Equation (\ref{9}) shows that the point-wise product $f\cdot g$ can also be interpreted as the $\star$-action 
of a differential operator $X^\star _f$ on $g$
\begin{equation}
f\cdot g = X_f^\star \star g = (X^\star _f \star g), \label{10}
\end{equation}
where
\begin{equation}
X^\star _f = \sum_{n=0}^\infty \frac{1}{n!}\Big(-\frac{i}{2}\Big)^n
\theta^{\rho_1\sigma_1}\dots\theta^{\rho_n\sigma_n}(\partial_{\rho_1}\dots\partial_{\rho_n}f)
\star \partial^\star _{\sigma_1}\dots\partial^\star _{\sigma_n} .\label{11}
\end{equation}
From the associativity of the $\star$-product follows immediately
\begin{equation}
f\cdot g\cdot h = X^\star _{f\cdot g} \star h = X^\star _f \star X^\star _g \star h .\label{12}
\end{equation}
The differential operators $X^\star _f$ represent the usual algebra of functions
\begin{equation}
X^\star _f \star X^\star _g = X^\star _{f\cdot g} .\label{13}
\end{equation}

\Section{Gauge transformations}
Ordinary gauge transformations are Lie algebra-valued
\begin{equation}
\alpha (x) = \alpha^a(x) T^a ,\quad [ T^a, T^b] = if^{abc}T^c\, .\label{14}
\end{equation}
The gauge transformation of a field is
\begin{equation}
\delta_\alpha \psi (x) = i\alpha(x)\psi(x) = i\alpha^a(x) T^a\psi(x) ,\label{15}
\end{equation}
i.e. $\delta_\alpha \psi =i\alpha\cdot\psi$.
This can be viewed as a $\star$-action 
\begin{equation}
\hat{\delta}_\alpha \psi = iX^\star_{\alpha^a}\star T^a\psi 
= iX^\star_\alpha \star \psi = i\alpha\cdot\psi  .\label{16}
\end{equation}
When we deal with a gauge theory in physics we not only use the  Lie algebra but also the corresponding Hopf algebra obtained from the comultiplication rule
\begin{eqnarray}
\Delta(\delta_\alpha) (\phi\otimes\psi) \hspace*{-2mm}&=&\hspace*{-2mm} 
(\delta_\alpha\phi )\otimes \psi + \phi \otimes (\delta_\alpha\psi ) ,\nonumber\\
\Delta(\delta_\alpha) \hspace*{-2mm}&=&\hspace*{-2mm} 
\delta_\alpha\otimes 1 + 1\otimes \delta_\alpha .\label{17}
\end{eqnarray}
The transformation of the product of fields is
\begin{equation}
\delta_\alpha (\phi\cdot\psi) = \delta_\alpha \mu\{\phi\otimes\psi\}
= \mu \Delta(\delta_\alpha) (\phi\otimes \psi) .\label{18}
\end{equation}
But there are different ways to extend a Lie algebra to a Hopf algebra. A convenient way is by a twist ${\cal F}$, that is a bilinear differential operator acting on a tensor product of functions. A well known example is
\begin{equation}
{\cal F} = e^{-\frac{i}{2}\theta^{\rho\sigma}\p_\rho\otimes \p_\sigma} .\label{19}
\end{equation}
It satisfies all the requirements for a twist \cite{Drinfeld:1990,Chari-Presley} and therefore gives rise to a new coproduct 
(the dual description of twisted gauge transformations was already introduced in \cite{Oeckl:2000eg}; see also \cite{Vassilevich:2006tc})
\begin{eqnarray}
&&\Delta_{\cal F}(\hat\delta_\alpha)(\phi\otimes \psi) = i{\cal F}(\alpha \otimes
1+1\otimes \alpha){\cal F}^{-1}(\phi\otimes \psi)\label{20} \\ 
&& \hspace{-0.4cm} =\sum_n(\frac{-i}{2})^n\frac{\theta^{\mu_1 \nu_1}\cdots\theta^{\mu_n \nu_n}}{n!}(\hat\delta_{\partial_{\mu_1}\cdots\partial_{\nu_n}\alpha}\otimes\partial_{\nu_1}\cdots\partial_{\nu_n}+\partial_{\mu_1}\cdots\partial_{\mu_n}\otimes\hat\delta_{\partial_{\nu_1}\cdots\partial_{\nu_n}\alpha})(\phi\otimes
\psi)\,.\nonumber
\end{eqnarray}
This coproduct defines a new Hopf algebra, the Lie algebra is extended by the derivatives, the comultiplication is deformed. 
This twist can also be used to deform  Poincar{\' e} transformations \cite{Oeckl:2000eg, Wess:2003da, Chaichian:2004za,Aschieri:2005yw} 
respectively diffeomorphisms \cite{Aschieri:2005yw,Aschieri:2005zs,Meyer:2005as}. In \cite{Balachandran:2006qg} gauge
theories consistent with twisted diffeomorphisms where constructed without deforming the coproduct for gauge transformations.

We now look at the transformation law of products of fields based on the deformed coproduct (\ref{20}).
\begin{equation}
\hat{\delta}_\alpha (\phi\star\psi) = \mu_\star \{\Delta_{\cal F}(\hat{\delta}_\alpha) (\phi\otimes \psi)\}, \label{21}
\end{equation}
where $\mu_\star$ is defined in (\ref{2}) and $\hat{\delta}_\alpha$ in (\ref{16}). We obtain
\begin{equation}
\hat{\delta}_\alpha (\phi\star\psi) 
= iX^\star_{\alpha^a}\star
\Big( ( T^a\phi )\star \psi + \phi\star (T^a\psi) \Big)  .\label{22}
\end{equation}
Note that the operator $X^\star_{\alpha^a}$ is at the left of both terms, this is due to the coproduct $\Delta_{\cal F}$. Formula (\ref{22}) is different from 
\begin{equation}
\hat{\delta}_\alpha (\phi\star\psi) = 
(\hat{\delta}_\alpha\phi )\star \psi + \phi \star (\hat{\delta}_\alpha\psi ) .\label{23}
\end{equation}
It is exactly the requirement that the $\star$-product of two fields should transform as (\ref{22}) that leads to the twist ${\cal F}$. It is by the twisted 
coproduct that the $\star$-product of fields transforms like (\ref{16}) again. The commutator of two gauge transformation closes
in the usual way 
\begin{equation}
\hat{\delta}_\alpha \hat{\delta}_\beta -\hat{\delta}_\beta \hat{\delta}_\alpha =\hat{\delta}_{-i[\alpha,\beta]} \,.
\end{equation} 

To construct an invariant Lagrangian we have to introduce covariant derivatives
\begin{equation}
D_\mu\psi = \p_\mu\psi - i A_\mu\star\psi . \label{24}
\end{equation}
From 
\begin{eqnarray}
\hat{\delta}_\alpha \psi = i X^\star _{\alpha^a}\star (T^a \psi) \nonumber
\end{eqnarray}
we find
\begin{equation}
\hat{\delta}_\alpha (D_\mu\psi) = iX^\star_{\alpha^a}\star \big( T^a(D_\mu\psi)\big)
\label{25}
\end{equation}
if we use the proper comultiplication for the term $A_\mu\star\psi$ in the covariant derivative and if the vector field transforms as follows
\begin{equation}
\hat{\delta}_\alpha A_\mu = \p_\mu \alpha + i X^\star_{\alpha^a}\star [T^a , A_\mu] . \label{26}
\end{equation}
This can also be written in the familiar way: 
\begin{equation}
\hat{\delta}_\alpha A_\mu = \p_\mu \alpha + i  [\alpha , A_\mu] . \label{27}
\end{equation}
The transformation would take Lie algebra-valued objects to Lie algebra-valued objects. For reasons that will become clear in the following we 
will assume the hermitian field $A_\mu$ to be $n\times n$ matrix valued where $n$ is the dimension of the Lie algebra representation.
Formula (\ref{26}) will still be true in that case.

The field-strength tensor can be obtained as usual
\begin{eqnarray}
F_{\mu\nu} \hspace*{-2mm}&=&\hspace*{-2mm} i[ D_\mu \ds D_\nu ] , \nonumber\\
\hspace*{-2mm}&=&\hspace*{-2mm} \partial _\mu A_\nu - \partial _\nu A_\mu 
-i [A_\mu \ds A_\nu ] .\label{28}
\end{eqnarray}
Using the deformed coproduct and the gauge variation of the potential
we derive the following transformation law,
\begin{eqnarray}
\hat{\delta}_\alpha F_{\mu\nu} \hspace*{-2mm}&=&\hspace*{-2mm}  
i X^\star_{\alpha^a}\star [T^a, F_{\mu\nu}] \label{29}\\
\hspace*{-2mm}&=&\hspace*{-2mm}i  [\alpha , F_{\mu\nu}] . \nonumber
\end{eqnarray}

\section{Field equations}

With the tensor $F_{\mu\nu}$ and the covariant derivatives we can construct invariant Lagrangians. Starting from 
the usual invariant Lagrangians we replace the point-wise  product by the $\star$-product and the 
comultiplication (\ref{18}) with (\ref{20}). We convince ourselves that we can construct an invariant Lagrangian 
under the deformed Hopf algebra. The expression $F^{\mu\nu}\star F_{\mu\nu}$ transforms as follows
\begin{eqnarray}
\hat{\delta}_\alpha (F^{\mu\nu}\star F_{\mu\nu}) \hspace*{-2mm}&=&\hspace*{-2mm}  
i X^\star_{\alpha^a}\star [T^a, F^{\mu\nu}\star F_{\mu\nu}] \label{30}\\
\hspace*{-2mm}&=&\hspace*{-2mm}i  [\alpha , F^{\mu\nu}\star F_{\mu\nu}] . \nonumber
\end{eqnarray}
This leads to an invariant and real action
\begin{equation}
S_g = c_1\int{\mbox{d}}^4 x\hspace{1mm} {\mbox{Tr}} (F^{\mu\nu}
\star F_{\mu\nu}) . \label{Sg}
\end{equation}
The integral introduced in (\ref{Sg}) has the trace property
\begin{equation}
\int{\mbox{d}}^4 x\hspace{1mm}(f\star g) = \int{\mbox{d}}^4 x\hspace{1mm}(f\cdot g)
= \int{\mbox{d}}^4 x\hspace{1mm}(g\star f) . \label{cyc}
\end{equation}
Therefore we obtain the field equations by writing the varied field to the very left. 
Varying  with respect to the matrix algebra-valued field $A_\mu$ leads to the field equations
\begin{equation}
\big( \partial_\mu F^{\mu\rho} \big) _{AB} - i \big( [ A_\mu \ds F^{\mu\rho} ] \big)_{AB} =0 .\label{eomA}
\end{equation}
Here $A$ and $B$ are matrix indices.

From the field equations and the antisymmetry of $F^{\mu\nu}$ in $\mu$ and $\nu$ follows the consistency requirement
\begin{equation}
\partial_\rho \Big( i [ A_\mu \ds F^{\mu\rho} ] \Big) =0 \label{divj} .
\end{equation}
To show (\ref{divj}) we have to use the equation of motion (\ref{eomA}). We calculate
\begin{equation}
\partial_\rho \Big( i [ A_\mu \ds F^{\mu\rho} ] \Big) 
= i [ \partial_\rho A_\mu \ds F^{\mu\rho} ] + i [ A_\mu \ds \partial _\rho F^{\mu\rho} ].
\label{divj1}
\end{equation}
In the second term we insert the field equation (\ref{eomA}). In the first term we complete $\partial_\rho A_\mu$ to the tensor $F_{\rho\mu}$ by adding and subtracting the respective terms. We then use
\begin{equation}
[ F_{\mu\rho} \ds F^{\mu\rho} ] = 0, \nonumber
\end{equation}
and obtain
\begin{eqnarray}
+ \frac{(i)^2}{2} \lb \lb A_\rho \ds A_\mu \rb \ds F^{\mu\rho} \rb
+ \frac{(i)^2}{2} \lb A_\mu \ds \lb A_\rho \ds F^{\mu\rho} \rb \rb
- \frac{(i)^2}{2} \lb A_\rho \ds \lb A_\mu \ds F^{\mu\rho} \rb \rb  = 0\nonumber
\end{eqnarray}
for the right hand side of equation (\ref{divj1}). That it vanishes follows from the Jacobi identity.
Thus, we obtained a conservation law
\begin{eqnarray}
J^\rho \hspace*{-2mm}&=&\hspace*{-2mm}   i [ A_\mu \ds F^{\mu\rho} ] ,\label{jg} \\
&& \hspace*{-1.2cm} \partial_\rho J^\rho = 0 .\nonumber
\end{eqnarray}

From (\ref{28}) follows that $F_{\mu\nu}$ is enveloping
algebra valued if $A_{\mu}$ is. From the field equation follows that
$A_{\mu}$ and $F_{\mu\nu}$ will remain enveloping algebra valued in the $n$-dimensional representation of the Lie algebra.
Thus, we try to replace matrix algebra valued by enveloping algebra
valued for $A_{\mu}$. 
As an example we treat the case $SU(2)$ in
the two-dimensional representation. In this representation the generators
$T^{a}$ of the Lie algebra satisfy the relations\begin{equation}
[T^{a},T^{b}]=i\epsilon^{abc}T^{c}\label{eq: Lie commutator}\end{equation}
 and \begin{equation}
\{ T^{a},T^{b}\}=\frac{1}{2}\delta^{ab}\,.\label{eq: Lie anti-comm}\end{equation}

Note that (\ref{eq: Lie commutator}) is valid for any representation. The anticommutator is representation dependent. Equation 
(\ref{eq: Lie anti-comm}) is only true in the two dimensional representation. In our example we can write $A_\mu$ as follows:
\[
A_{\mu}=B_{\mu}+A_{\mu}^{d}T^{d}\,.\]
 This is consistent with the gauge transformations; the field equations
are a consequence of (\ref{eq: Lie commutator}) and (\ref{eq: Lie anti-comm}). 

The tensor $F_{\mu\nu}$ is easy to calculate following (\ref{28}):\[
F_{\mu\nu}=G_{\mu\nu}+\tilde{F}_{\mu\nu}^{d}T^{d}\,,\]
 where \begin{eqnarray}
G_{\mu\nu} & = & \partial_{\mu}B_{\nu}-\partial_{\nu}B_{\mu}-i[B_{\mu}\stackrel{\star}{,}B_{\nu}]-\frac{i}{4}[A_{\mu}^{d}\stackrel{\star}{,}A_{\nu}^{d}]
\nonumber \\
\tilde{F}_{\mu\nu}^{d} & = &
\partial_{\mu}A_{\nu}^{d}-\partial_{\mu}A_{\nu}^{d}-i[B_{\mu}\stackrel{\star}{,}A_{\nu}^{d}]-i[A_{\mu}^{d}\stackrel{\star}{,}B_{\nu}]+\frac{1}{2}\{
A_{\mu}^a\stackrel{\star}{,}A_{\nu}^b\}\epsilon^{abd}\,. \label{EOM}
\end{eqnarray}
 Varying the Lagrangian (\ref{Sg}) with respect to $B_{\mu}$ and
$A_{\mu}^{d}$ leads to the field equations 
\begin{eqnarray}
\partial^{\mu}G_{\mu\nu}-i[B^{\mu}\stackrel{\star}{,}G_{\mu\nu}]-\frac{i}{4}[A^{\mu a}\stackrel{\star}{,}\tilde{F}_{\mu\nu}^{a}] & = & 0\nonumber \\
\partial^{\mu}\tilde{F}_{\mu\nu}^{d}-i[A^{\mu d}\stackrel{\star}{,}G_{\mu\nu}]-i[B^{\mu}\stackrel{\star}{,}\tilde{F}_{\mu\nu}^{d}]+\frac{1}{2}\epsilon^{abd}\{ A_{\mu}^{a}\stackrel{\star}{,}\tilde{F}_{\mu\nu}^{b}\} & = & 0\,.\label{eq:Field equn}
\end{eqnarray}
These field equations are consistent. They describe a triplet of vector
fields $A_{\mu}^{d}$ as expected and a singlet $B_{\mu}.$ In the
limit $\theta\rightarrow0$, $B_{\mu}$ becomes a free field; it interacts
only via $\theta$ and higher order terms in $\theta$. The triplet
$A_{\mu}^{d}$ satisfies the usual field equations of  $SU(2)$ 
gauge theory in the limit $\theta \rightarrow 0$. For $\theta\neq0$ both the triplet and the singlet
couple to conserved currents but the current of $B_{\mu}$ has no $\theta$-independent
term.

We discover that the field equations (\ref{EOM}) with four conserved currents also have a larger symmetry structure, i.e. the gauge
transformations (\ref{14}) can also be enveloping algebra valued.   

\section*{Acknowledgements}

The authors would like to thank B. Jurco and M. Wohlgenannt for
helpful 
discussions and comments. F.M. would like to thank L.
\'Alvarez-Gaum\'e and M. V\'azquez-Mozo for many 
useful discussions. M.D. would like to thank the Centre for
Mathematical 
Physics, DESY and University of Hamburg, for hospitality.
Financial support from the DFG program (1096) "Stringtheorie im
Kontext von 
Teilchenphysik", the EC contract 
MRTN-CT-2004-005104, the italian MIUR contract
PRIN-2005023102 and the
Max-Planck Institute for Physics Munich is gratefully
acknowledged.

\appendix 

\section{Notations on coproduct}
We might rewrite the first part of Section 3 (up to formula (\ref{21})) 
using a 
more mathematically oriented language. We consider (the 
semidirect product of) the Lie algebra of
the local gauge group $\alpha(x)=\alpha(x)^aT^a$ and of 
translations. 
The undeformed coproduct on the generators is
\begin{eqnarray}
\Delta(\alpha) \hspace*{-2mm}&=&\hspace*{-2mm} 
\alpha\otimes 1 + 1\otimes \alpha=\al_1\otimes\alpha_2 \\
\Delta(\partial_\nu) \hspace*{-2mm}&=&\hspace*{-2mm} 
\partial_\nu\otimes 1 + 1\otimes \partial_\nu
\end{eqnarray}
where $\alpha_1\otimes\alpha_2$ 
is a convenient notation for
$\alpha\otimes 1 + 1\otimes \alpha$.
In  $\al_1\otimes\alpha_2$ a sum is understood 
so that $\alpha_1$ respectively assumes the  values
$\alpha$ and 1 (and similarly for $\alpha_2$). 

The action of $\al$ and $\partial_\nu$ on fields is given by the 
gauge transformation $\delta_\al\phi$ and by the usual derivative action
$\partial_\nu\phi$. From the coproduct $\Delta$ we have the action of $\al$ 
on the product of fields
\begin{equation}
\delta_\alpha (\phi\cdot\psi) = \delta_\alpha \mu\{\phi\otimes\psi\}
= \mu\{ (\delta_{\alpha_1}\phi)\otimes 
 (\delta_{\alpha_2}\psi) \}
=(\delta_{\alpha_1}\phi) 
 (\delta_{\alpha_2}\psi)\,,
\end{equation}
and the usual Leibniz rule for partial derivatives. 

We now deform the coproduct $\Delta$ by using the twist ${\cal F}$,
and obtain the new coproduct 
\begin{eqnarray}
\Delta_{\cal F}(\alpha) &=& {\cal F}(\alpha \otimes
1+1\otimes \alpha){\cal F}^{-1}\\ \label{20mat}
&=&\sum_n\left(\frac{-i}{2}\right)^{n}\frac{\theta^{\mu_1\nu_1}\cdots\theta^{\mu_n \nu_n}}{n!}({\partial_{\mu_1}\cdots\partial_{\mu_n}\alpha}\otimes\partial_{\nu_1}\cdots\partial_{\nu_n} \nonumber \\
&&\,\,\,\,\,\,\,\,\,\,\,\,\,\,\,\,\,\,\,\,\,\,\,\,\,\,\,\,\,\,\,\,\,\,\,\,\,\,\,\,\,\,\,\,\,\,\,\,\,\,\,\,\,
\,\,\,\,\,\,\,\,\,\,\,\,\,\,\,\,\,\,\,
 +\partial_{\mu_1}\cdots\partial_{\mu_n}\otimes
{\partial_{\nu_1}\cdots\partial_{\nu_n}\alpha})\nonumber\\ 
&=&\alpha_{1_{\cal F}}\otimes \alpha_{2_{\cal F}}
\,\nonumber
\end{eqnarray}
where  in the convenient notation
$\alpha_{1_{\cal F}}\otimes \alpha_{2_{\cal F}}$   sum over $n$ 
is understood.

Since ${\cal F}$ satisfies all the requirements for a twist, we obtain 
that the universal enveloping algebra generated by derivatives
$\partial_\nu$ and gauge parameters $\al(x)=\al(x)^aT^a$ has been equipped
with a new coproduct, the twisted coproduct $\Delta_{\cal F}$. This
coproduct defines a new Hopf algebra.

We now consider the noncommutative action of $\al$ and $\partial_\nu$ 
on fields and on $\star$-products of fields. The noncommutative action
of partial derivatives is undeformed, see (\ref{4'}). 
Our noncommutative gauge
principle is implemented by defining 
noncommutative gauge transformations as
\begin{equation}
\hat{\delta}_\alpha \psi = iX^\star_{\alpha^a}\star T^a\psi 
= iX^\star_\alpha \star \psi = i\alpha\cdot\psi  \,
\end{equation}
and 
\begin{eqnarray}
\hat{\delta}_\alpha (\phi\star\psi) 
&=& 
\mu_\star \{(\hat{\delta}_{\alpha_{1_{\cal F}}} \phi)\otimes
(\hat{\delta}_{\alpha_{2_{\cal F}}} \psi)\}, \label{21mat} \nonumber
\\ 
&=&
\mu_\star\big\{\sum_n\left(\frac{-i}{2}\right)^{n}\frac{\theta^{\mu_1 \nu_1}\cdots\theta^{\mu_n \nu_n}}{n!}({\partial_{\mu_1}\cdots\partial_{\mu_n}\alpha})\phi\otimes(\partial_{\nu_1}\cdots\partial_{\nu_n}\psi)
\nonumber \\
& & \,\,\,\,\,\,\,\,\,\,\,\,\,\,\,\,\,\,\,\,\,\,\,\,\,\,\,\,\,\,\,\,\,\,\,\,\,\,\,\,\,\,\,\,\,\,\,\,\,\,\,\,\,
\,\,\,\,\,\,\,\,\,\,\,\,\,\,\,\,\,\,\,+
 (\partial_{\mu_1}\cdots\partial_{\mu_n}\phi)\otimes
({\partial_{\nu_1}\cdots\partial_{\nu_n}\alpha})\psi
\big\}
\nonumber\\
&=&
 iX^\star_{\alpha^a}\star
\Big( ( T^a\phi )\star \psi + \phi\star (T^a\psi) \Big)  \,.
\end{eqnarray}
where $\mu_\star$ is defined in (\ref{2}).
The Hopf algebra structure obtained with the twisted coproduct
$\Delta_{\cal F}$ insures the consistency of the 
noncommutative gauge transformation $\hat\delta_\al$
on $\star$-products of fields.

\bibliographystyle{diss}
\bibliography{literature_gravity}

\end{document}